# Feasibility to detect rapid change and disappearance of seagrass: Lessons from nearly 80 years of vegetation change in the Ako, Seto Inland Sea, Japan

Takehisa Yamakita[1,2], Yoji Igarashi[1], Akira Eto[1,3], Ken Ishida[1], Masaaki Iiyama[4]

[1]Japan Agency for Marine-Earth Science and Technology (JAMSTEC), Application Laboratory (APL), 3173-25, Showa-machi, Kanazawa-ku, Yokohama, Kanagawa, Japan – Yamakitat@jamstec.go.jp
[2] The University of Tokyo, Graduate School of Agricultural and Life Sciences, 1-1-1 Yayoi, Bunkyo City, Tokyo, Japan
[3] Tokyo University of Marine Science and Technology, 4-5-7 Kohnan, Minato-ku, Tokyo, Japan
[4] Shiga University, Faculty of Data Science, 1-1-1 Banba, Hikone City, Shiga, Japan

**Keywords:** Aquatic Plants, Essential Biodiversity Variables (EBVs), Global warming, Marine heatwave, Satellite Constellation, AI Image segmentation.

**Abstract**

Seagrass meadows provide essential ecosystem services but are declining globally, creating an urgent need for continuous monitoring and management. However, the high cost of commercial satellite imagery and field surveys limits sustained observation, highlighting the need for practical workflows that integrate historical aerial photographs with low-cost or freely available satellite data using open-source methods. This study analyses the Ako tidal flat in the Seto Inland Sea, Japan, where nearly all *Zostera marina* disappeared within a single year in 2025. Using aerial photographs from the 1940s onward, high-resolution satellite imagery, GRUS images (2.5–5 m), and monthly Sentinel-2 composites (10 m), we reconstructed approximately 80 years of seagrass distribution. YOLO-based segmentation using deep learning achieved high accuracy (overall accuracy ≥ 0.9) across these datasets; although species could not be discriminated, the models captured the major temporal dynamics in vegetation area. The long-term mean seagrass area was 6.8 ha, but values fluctuated widely, from 3.5 ha in 1974 to 41.3 ha in 1989 except 0.2 ha in 2025. Sentinel-2 composites from 2019 to 2026 revealed clear seasonality, with vegetation increasing in early summer and declining from autumn. In 2025, however, the area decreased sharply after summer and remained anomalously low throughout the winter of 2025–2026. Our results, indicating that the 2025 event was not a normal fluctuation but a rapid ecosystem shift involving the loss of the dominant canopy-forming species, most plausibly driven by regionally elevated summer water temperatures. The findings also have implications for seagrass Essential Ocean Variables (EOVs) and the State of Nature (SoN) metrics used in TNFD-aligned nature-related disclosures. Unlike forests, seagrass meadows require finer temporal resolution because both pronounced seasonality and abrupt collapse strongly influence area-based indicators. Therefore, in addition to previously noted issues such as species-level classification accuracy, we recommend that (1) baselines be defined over the longest available record and justified ecologically, (2) seasonal standardization be applied before inter-annual comparisons, and (3) years with extreme area anomalies be flagged rather than used as reference points.

## 1. Introduction

Seagrasses are flowering plants (Angiosperms) that have secondarily colonised marine environments, analogous to the evolutionary return of whales to the sea. Seagrass ecosystems provide essential functions and ecosystem services to coastal areas by introducing structural complexity through plant tissue in intertidal zones. These services include habitats for small intertidal organisms, nursery and feeding grounds for fishery species such as squid spawning aggregations, shoreline protection through sediment stabilization, and water purification. Recently, seagrasses have been gaining broad attention as a major component of blue carbon. In Japan in particular—where the world's first voluntary blue carbon credits targeting seagrass beds and macroalgal beds were issued—private-sector restoration activities have been increasing (Yamakita, 2025; Kuwae et al., 2026; Fig.1). Yet many seagrass beds worldwide are in decline or unknown status (Waycott et al., 2009; McKenzie et al., 2020). Adaptive management of these ecosystems requires the ability to detect both long-term trends (Yamakita et al., 2011) and abrupt collapses across spatial and temporal scales relevant to policy and local management.

Continuous monitoring of seagrass faces practical limitations. High-resolution commercial imagery and intensive field surveys provide high accuracy, but they tend to be prohibitively expensive, particularly when long-term time-series data or extensive coastal areas are involved (Duffy et al., 2025). On the other hand, combining historical aerial photographs and recently available low-cost or free satellite data with open-source software (OSS) has the potential to enable accessible monitoring by local citizens, experts from other fields, or even countries and organizations with limited budgets. However, this approach remains limited to case studies and has not yet been utilized as an integrated monitoring method.

This study focuses on an eelgrass (*Zostera marina*) bed in Ako, eastern Seto Inland Sea, Japan, where nearly all *Z. marina* disappeared within a single year, leaving some patches of the smaller seagrass *Zostera japonica* and green algae. We present a case study demonstrating that the combination of high-resolution historical aerial photographs and recent low-cost or freely available satellite data, processed primarily using free and open-source software (including FOSS4G), can detect both long-term trajectories and an acute ecosystem collapse.

This project was conducted in alignment with the framework of the Asia-Pacific Marine Biodiversity Observation Network (APMBON) (Takeuchi et al., 2021). We also discuss how such observations can contribute to seagrass Essential Ocean Variables (EOVs) / Essential Biodiversity Variables (EBVs) (Duffy et al., 2025) as defined under the Marine Biodiversity Observation Network (MBON) / Biodiversity Observation Network (GEO BON), and aligned with the indicators of the Kunming-Montréal Global Biodiversity Framework (KMGBF), and the State of Nature Metrics (SoN) assessment proposed by the Nature Positive Initiative (NPI) in the context of the Taskforce on Nature-related Financial Disclosures (TNFD) and/or the management of Other Effective area-based Conservation Measures (OECMs).

## 2. Methods

### 2.1 Study site

The *Zostera marina* seagrass beds of the Seto Inland Sea are reported to have declined by more than 70% from their historical extent. We selected the tidal flat of the Ako coast as our study site (Figure 1). This site is located adjacent to the Hinase area, where the most active seagrass restoration efforts in the Seto Inland Sea are currently undertaken (Hori and Sato, 2021; Tsurita et al., 2018). Land transformation from saltpans to industrial land and public parks is also a typical form of land-use change in this region. Although some restoration activities have also been conducted at this site recently, it is considered suitable for observing near natural-state vegetation dynamics that are free from intensive human intervention.

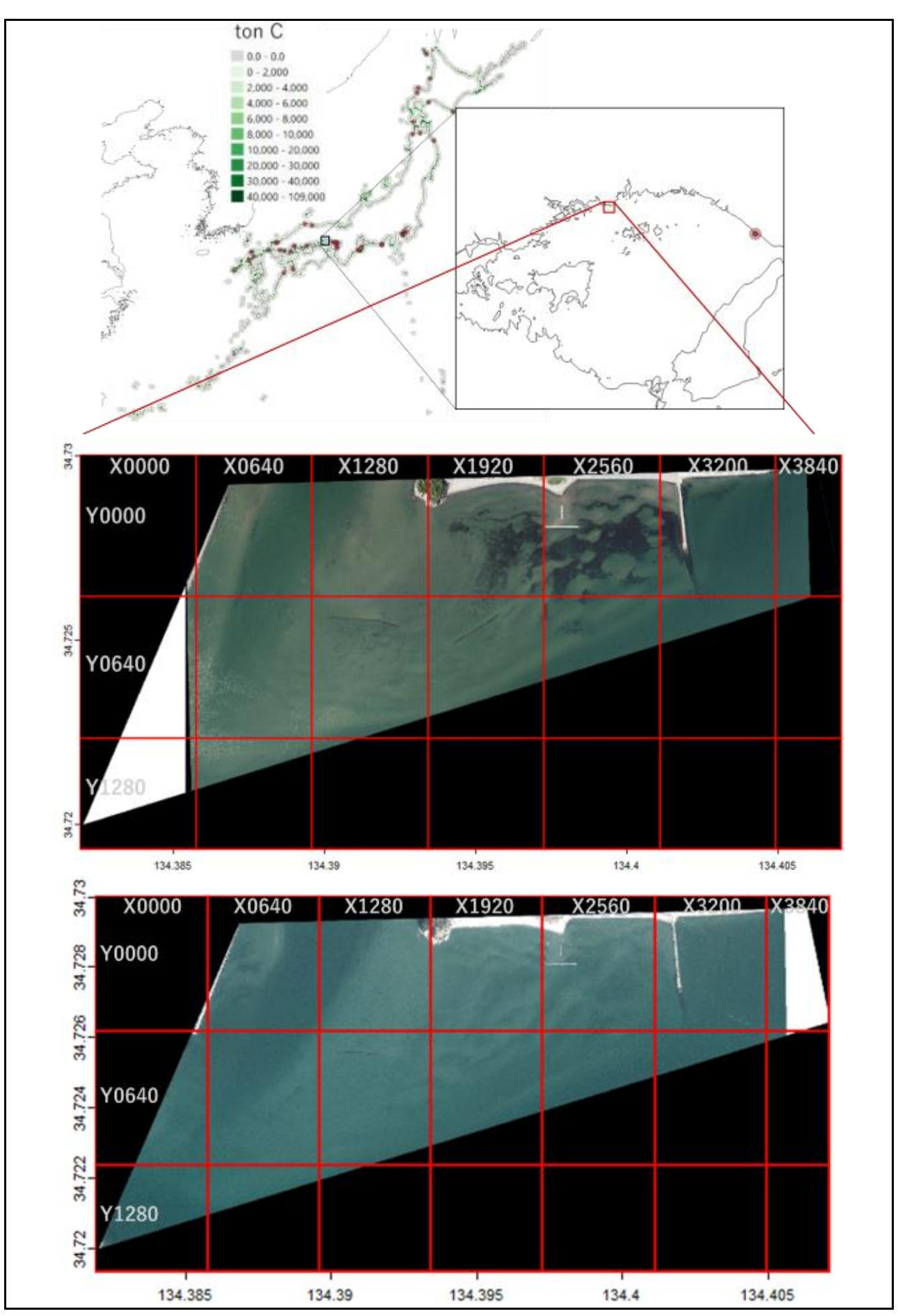

Figure 1. Study site of the Ako coast and the analysis grids used in this study. The upper image shows the estimated carbon absorption (Yamakita 2022; Saito et al., 2021) and location of the Blue Carbon Credit project (JBE website; Yamakita 2025; Kuwae et al., 2026). The lower images were taken by GSI in 2010 and 2025. Coastline based on Wessel and Smith (1996).

### 2.2 Data and Analysis methods for long term change

We compiled historical aerial photographs taken by the Geospatial Information Authority of Japan (GSI) and the US military dating back to the 1940s to reconstruct past seagrass distribution (Table 1). We additionally incorporated sub-meter-scale high-resolution satellite data using the WorldView series provided by the Secure Watch Service.

After geometric correction and/or orthorectification using QGIS and Geoplotter (Sakamoto et al., 2000; Asia Air Survey Co., Ltd.), seagrass polygons were extracted from historical images using a YOLOv8 segmentation model (Jocher et al., 2023), extending the earlier workflow reported by Yamakita et al. (2019), which also compared deep learning models for RGB color and black-and-white (B/W) grayscale images (Tahara et al., 2022).

| Type | Color | Year | MM | Scale/ resolution |
|---|---|---|---|---|
| Airplane US | B/W | 1947 | 8 | 43799 |
| Airplane GSI | B/W | 1966 | 11 | 20000 |
| Airplane GSI | RGB | 1975 | 2 | 8000 |
| Airplane GSI | B/W | 1975 | 1 | 40000 |
| Airplane GSI | RGB | 1980 | 9 | 10000 |
| Airplane GSI | B/W | 1989 | 4 | 40000 |
| Airplane GSI | RGB | 1999 | 4 | 30000 |
| Airplane GSI | RGB | 2010 | 6 | 10000 |
| WorldViewII | RGB | 2019 | 11 | 50cm |
| WorldViewII | RGB | 2023 | 2 | 48cm |
| Airplane GSI | RGB | 2025 | 2 | 10000 |

Table 1. Historical aerial photographs and high-resolution satellite images used for long-term analysis.

We prepared the training data through visual interpretation and the most recent maps were validated by field observations including a drone survey. For the training data, we used information from labels created for the entire extent of two images taken in 2010 and 1999. Additionally, label information for a single grid was added from the images of each year. To ensure consistency between the training and validation data across images, the training data and the independent validation data were fixed to a single grid on the shore (X2560 and X3200 of Y0, respectively). Furthermore, to ensure consistency in image analysis, images with different resolutions were upscaled/downscaled to the same resolution and automatic color match was applied to all images before conducting the deep-learning analysis.

### 2.3 Data used for recent frequent monitoring

For contemporary seasonal dynamics, we used imagery from optical satellite constellations with 5 to 10 meters resolution, including affordable commercial constellations, and freely available open missions. Images were acquired more than twice per year and we applied the same workflow used for classical aerial images (Table 2 and 3).

GRUS, operated by Axelspace Hold. Corp., was first launched in December 2018. It has a resolution of 2.5 meters for monochrome imagery and 5 meters for multispectral imagery, and it consists of a five-satellite constellation, providing approximately two passes per day. We selected appropriate images approximately two seasons per year from their data platform AxcelGlobe and used images after precise rectification.

Sentinel-2 data are part of the Copernicus Earth observation program led by the European Space Agency (ESA). It consists

of a constellation of three satellites, the first of which was launched in June 2015. Visible-band data have a resolution of 10 meters and can be acquired at five-day intervals. After verifying coverage and quality of data using the Sentinel-2 Image Downloader plugin for QGIS, we automatically extracted monthly median images with less than 20% cloud cover from Google Earth Engine. From 2019 to 2026, 62 monthly composites were successfully obtained, while 27 months yielded no valid imagery due to excessive cloud cover (Table 3.).

As with the historical images, we performed automatic classification by creating a YOLO segmentation model using these data. For the training and validation datasets, we provided two images (Feb. 2023 and Nov. 2023) with full-area labels of the vegetation, which were manually produced using high resolution images as reference (Wabnitz et al., 2008). We also added additional single-grid training data from five years, collected from the same locations of the historical images. For the additional training data, we used all obtained GRUS images except September 2025 and used the 2019-01,2019-11,2020-03,2021-03, and 2024-08 images for the Sentinel-2.

| Platform | Year | MM | DD |
|---|---|---|---|
| GRUS-E | 2022 | 04 | 12 |
| GRUS-B | 2022 | 10 | 16 |
| GRUS-C | 2023 | 02 | 12 |
| GRUS-C | 2023 | 11 | 08 |
| GRUS-E | 2024 | 03 | 18 |
| GRUS-B | 2024 | 06 | 12 |
| GRUS-B | 2025 | 1 | 3 |
| GRUS-A | 2025 | 9 | 1 |

Table 2. GRUS images used for recent seasonal analysis.

| Month / Year | 1 | 2 | 3 | 4 | 5 | 6 | 7 | 8 | 9 | 10 | 11 | 12 |
|---|---|---|---|---|---|---|---|---|---|---|---|---|
| 2019 | 3 | 0 | 1 | 2 | 2 | 1 | 0 | 2 | 2 | 0 | 3 | 1 |
| 2020 | 1 | 0 | 2 | 1 | 2 | 2 | 0 | 3 | 0 | 2 | 2 | 0 |
| 2021 | 2 | 2 | 3 | 0 | 2 | 0 | 0 | 0 | 2 | 3 | 1 | 1 |
| 2022 | 2 | 0 | 0 | 2 | 2 | 3 | 1 | 0 | 0 | 2 | 2 | 2 |
| 2023 | 1 | 2 | 3 | 3 | 1 | 1 | 2 | 2 | 0 | 1 | 1 | 2 |
| 2024 | 0 | 0 | 1 | 0 | 1 | 0 | 0 | 2 | 2 | 1 | 2 | 0 |
| 2025 | 2 | 0 | 1 | 2 | 1 | 0 | 0 | 2 | 0 | 3 | 1 | 3 |
| 2026 | 5 | 2 | 1 | 0 | 1 | - | - | - | - | - | - | - |

Table 3. Number of Sentinel-2 scenes used to generate each monthly composite.

## 3. Results

### 3.1 Long-term changes in seagrass using aerial photographs

The total detected vegetation area across all grids showed marked temporal variation over the study period (1947–2025; results from B/W images are used here), with a coefficient of variation (CV) of 1.09 across all 11 observation years, reflecting

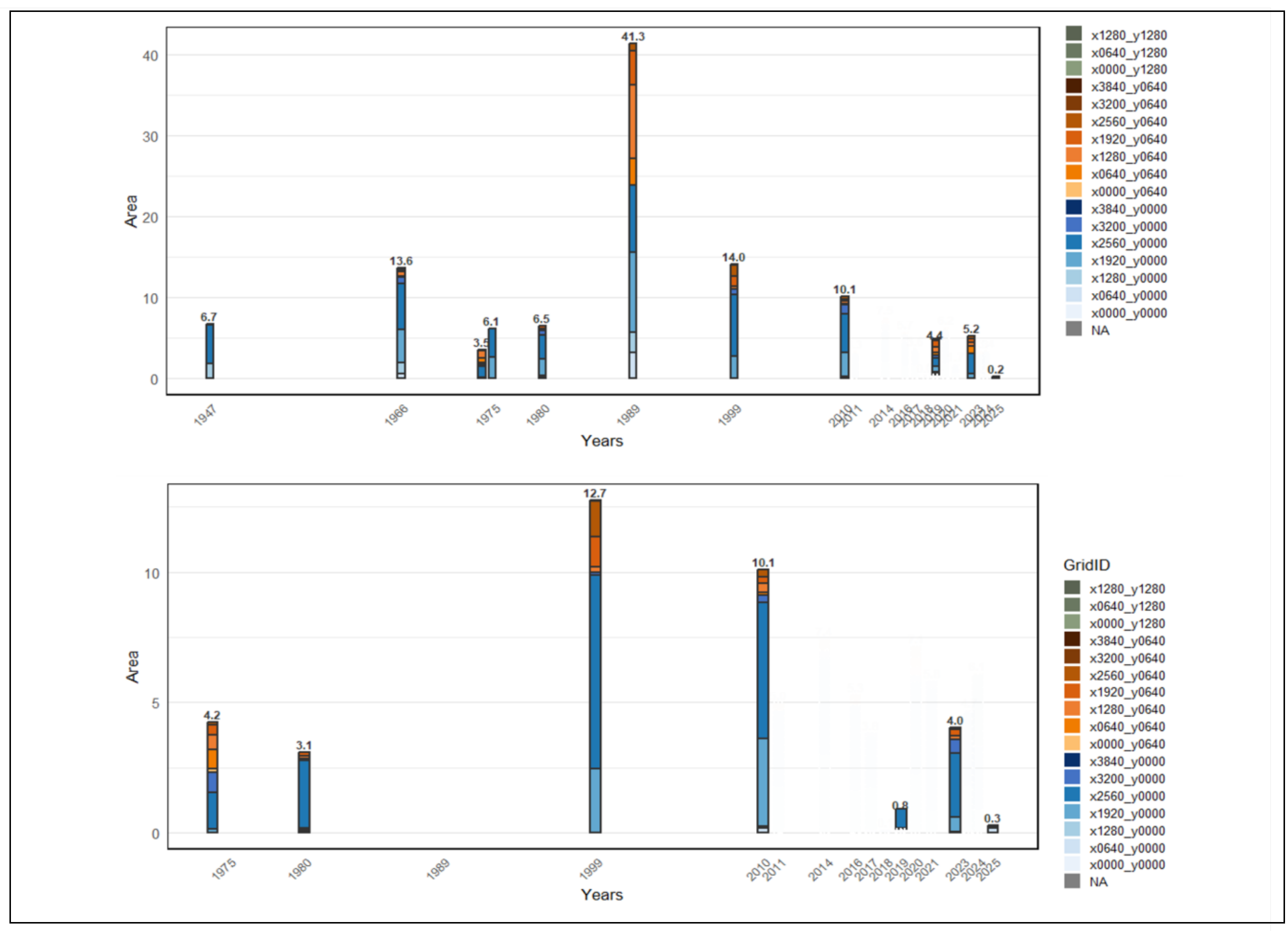

Figure 2. Temporal change of vegetation area (ha) obtained from historical high-resolution aerial photographs and satellite imagery. (Upper: Black and white images since 1940s; Bottom: color images since 1974)

extreme inter-annual variability (Figure 2). The highest total area was recorded in 1989 (41.3 ha), followed by 1999 (14.0 ha) and 1966 (13.6 ha), indicating that vegetation was most extensively distributed around the 1990s. In contrast, detections in 1974 and 2019 showed considerably smaller values, with the total area declining to 3.5 ha in 1974, 4.5 ha in 2019, and reaching a minimum of 0.2 ha in 2025. The CV for the most recent decade (1989–2025) dropped to 0.79, suggesting that while year-to-year variability persisted, the range of fluctuation has narrowed substantially compared to the full record.

Within the most dominant shallow grid (x2560_y0000), the detected area peaked at 8.3 ha in 1989 and remained at high levels through 1966 (8.8 ha) and 1999 (7.6 ha) except in 1974 (1.3 ha). Some decline was observed from the 2000s, but the drop to 1.0 ha in 2019 is likely due to the condition of the image and tidal level. Based on eleven images on Google Earth from 2011 to 2014, the vegetation extent appears more consistent with the detected area in 2023 (2.4 ha). In 2025, the vegetation area in this grid had fallen to 0.06 ha, the lowest recorded value in the entire time series, suggesting a long-term reduction trend with fluctuations and a sudden disappearance.

When YOLO was applied to B/W images, the overall pixel-wise accuracy across all years with test data reached 0.959 as overall accuracy, with recall of 0.804, precision of 0.614, specificity of 0.969, and Cohen's kappa of 0.675. These results suggest that the model reliably excluded non-eelgrass areas, but some seagrass patches, especially fragmented or spectrally ambiguous areas, were still omitted. For RGB color images, the overall accuracy reached 0.965, with a seagrass recall of 0.811, precision of 0.663, background specificity of 0.975, and Cohen's kappa of 0.711.

## 3.2 Seasonal changes using Satellite Constellation

In the Sentinel-2 monthly data (2019–2026), the total area of vegetation showed significant fluctuations ranging from 0.2 to 20.5 ha (Figure 3; in November 2020 and June 2022). A distinct seasonal pattern was observed, with peaks occurring annually from May to July (monthly averages for the entire period: June: 16.0 ha, July: 17.6 ha), followed by a repeated pattern of rapid decline from late summer through autumn. However, observations for August and September were limited in many years due to cloud cover. Since vegetation in the offshore areas beyond y=640 is not sufficiently visible in aerial photographs—which are rarely taken during summer—and field surveys could not be conducted there, it remains unclear whether this pattern reflects seasonal fluctuations in seagrass or the presence of green or other algae.

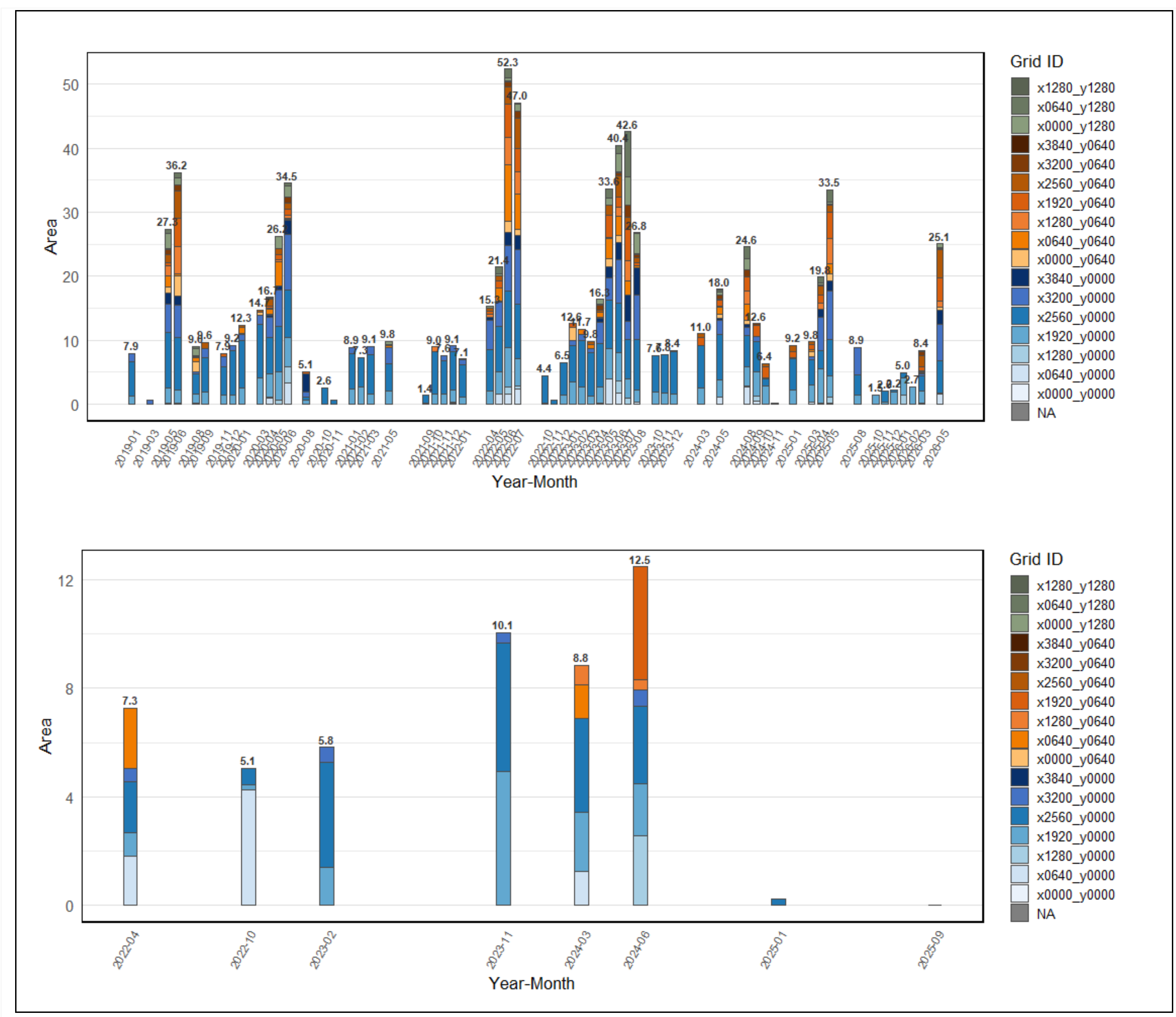

Figure 3. Seasonal changes in vegetation area (ha) obtained from satellite constellations. (Upper: Sentinel-2; Lower: GRUS)

In 2025, the detected area remained comparable to the 2019–2024 average (8.9 ha in May) until spring (13.2 ha in May, 2025), then decreased sharply from August, continuing at extremely low levels in October (0.6 ha). Even after entering 2026, these low levels continued until February (1.0 ha). Winter-season (Nov.–Jan.) values in 2025–2026 (1.5 ha SD=1.1) are well below the 2019–2024 winter mean (3.1 ha SD=1.5), at less than half of that mean, though still within the historical range of variability.

In terms of spatial patterns by grid, the three grids closest to the shore—x2560_y0000, followed by x1920_y0000 and x3200_y0000—showed high values, but since August 2025, the area in these major grids has also decreased significantly.

When compared with field surveys, this corresponds to vegetation where the smaller species *Zostera japonica* and green algae remain in some areas following the disappearance of the larger species *Zostera marina* (Figure 4.)

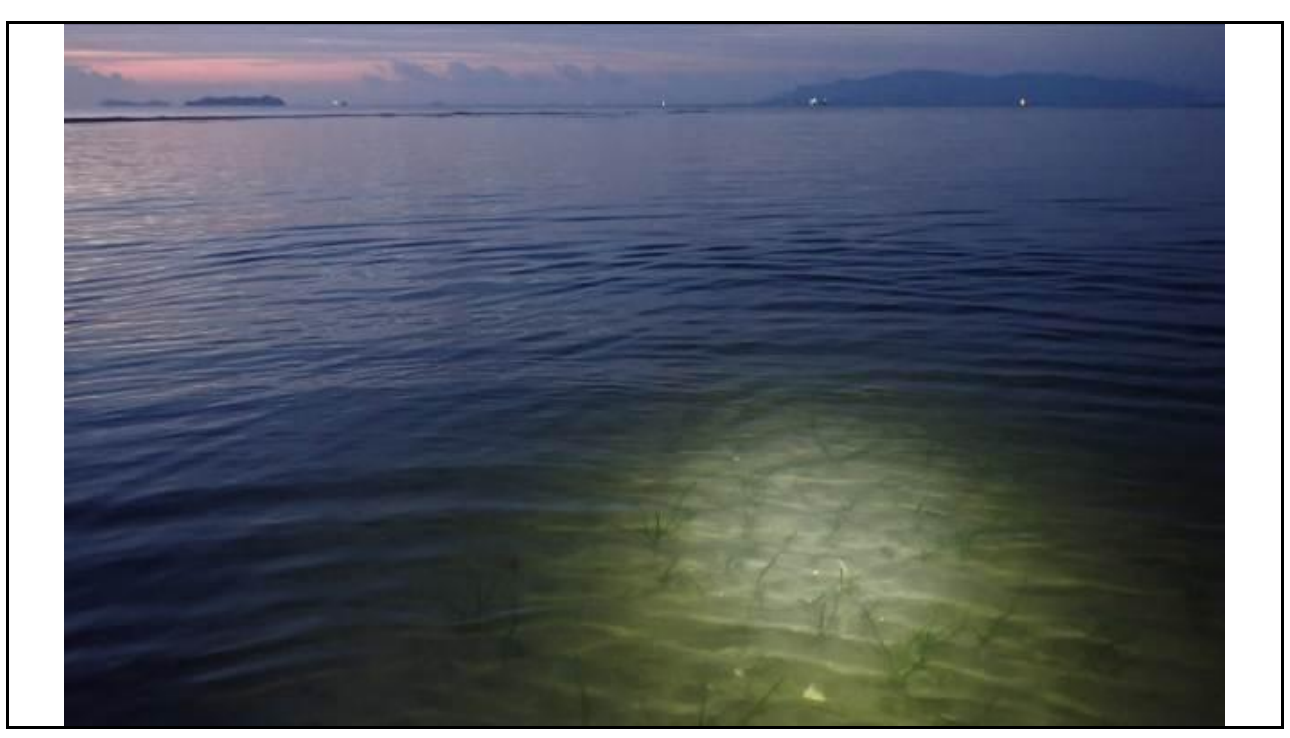

Figure 4. Residual eelgrass patch observed nearshore in the northern part of the x2560_y0000 grid on December 18, 2024

For the results using Sentinel-2 data, the overall accuracy reached 0.952, with a seagrass recall of 0.692, precision of 0.522, background specificity of 0.966, and Cohen's kappa of 0.570. The confusion matrix shows that 95% of background pixels are correctly identified as background, whereas approximately 69% of seagrass pixels are detected, with omission errors on the order of 31%. Although the proportion of other species such as *Zostera japonica* and green algae (*Ulva* spp.) within this vegetation is unknown, the results suggest that even 10-meter-resolution satellite imagery is sufficiently capable of tracking long-term changes in the area of seagrass bed vegetation.

The GRUS constellation results peaked at 12.5 ha in June 2024 and dropped to 0.0 ha by September 2025, with a small residual area still present in early 2025 (0.2 ha in January 2025) before disappearing thereafter. The mean area also differed markedly between periods, being 8.3 ha (SD = 2.8) up to 2025 and 0.1 ha (SD = 0.2) thereafter.

For the GRUS data, overall accuracy was 0.897, recall 0.575, precision 0.196, and Cohen's kappa 0.250. It should be noted that classifier models trained on color-corrected images performed significantly worse, so the final model was trained on uncorrected images but applied to color-corrected data for detection, to accommodate potential noise in GRUS imagery, especially in the early scenes acquired soon after launch. To compare GRUS-derived area estimates with those from Sentinel-2, we correlated the GRUS results with a moving average of Sentinel-2 estimates for the corresponding month and the preceding and following months; when no Sentinel-2 data were available for the same month, only the mean of the two neighboring months was used. This yielded moderate to high correlations: $R^2$ =0.67, p=0.07 for the whole area, and $R^2$=0.85, p<0.01 for grid x2560_y0000, the densest nearshore part of the site.

## 4. Discussion

### 4.1 Long-term changes of seagrass

The historical trajectory reconstructed in this study indicates that the seagrass community currently present is not simply a remnant of a stable, pre-industrial habitat; rather, it is a secondary vegetation type that expanded following large-scale coastal modification. Over almost eight decades, seagrass distribution underwent repeated phases of expansion and contraction. Most of the vegetation emerged after saltpans were discontinued in the 1960s, expanding until the late 1980s. During this phase the construction of offshore breakwaters attenuated wave energy and reduced turbidity. The vegetation subsequently decreased again until late 2010s. This trajectory suggests that both anthropogenic disturbance and natural variability have acted in both positive and negative ways on the system over time, highlighting the importance of historical baselines for interpreting contemporary change.

The decrease observed in 2025 does not necessarily exceed the historical amplitude of area change. The most recent decline in 2025 is distinguished not only by its magnitude alone, but also by two additional features. The decline was also distinguished by the extremely short timescale over which it occurred, and by the fact that *Z. marina* disappeared entirely rather than undergoing a gradual decrease. The complete loss of the dominant species indicates that the 2025 collapse reached a qualitative threshold—a probable tipping point in the ecological sense—rather than representing a point within normal interannual variability.

Despite the absence of direct environmental measurements at the study site, the most plausible proximate driver is anomalously high water temperature during the summer to early autumn of 2025. This corresponds with a well-documented mechanism for *Z. marina* under thermal stress (Nakamura et al., 2005; Marbà et al., 2022). This interpretation is consistent with a broader pattern of marine heatwave-associated ecosystem change in the Seto Inland Sea in 2025, including the decline of seagrass on the Eigashima Coast in 2024 (located in the same Harima-nada region of the Seto Inland Sea) and mass oyster mortality in 2025 around Hiroshima (on the west side of the Seto Inland Sea; Tanaka et al., 2025; Fishery Agency, 2026). Moreover, the concurrent rapid decline of *Z. marina* was also recorded at the Futtsu tidal flat, which is the largest eelgrass bed in Tokyo Bay in 2025 (Yamakita et al., under review). The co-occurrence of acute seagrass losses at multiple geographically separate sites strongly suggests a regional-scale climate driver rather than a site-specific perturbation.

Field surveys conducted in 2025 found rhizomes of *Z. marina* below the surface in areas accessible on foot, but no living shoots. *Z. japonica* patches remained present. This finding confirms that the 2025 decline represents lethal mortality of the dominant canopy species rather than temporary dormancy, and that satellite-detected residual area after August 2025 most likely reflects *Z. japonica* and *Ulva* spp. rather than recovering *Z. marina*.

Distinguishing between decline and disappearance is crucial for monitoring and management. Although high-resolution imagery does not allow for precise species identification, it enables detailed confirmation that most patches have disappeared in areas where eelgrass was previously observed. However, while conventional long-term trend analyses may detect the scale of change, they may not be able to identify the rate of decline or the potential for recovery.

### 4.2 Recent seasonal changes of seagrass

The seasonal satellite time series reveals that the vegetation, mainly seagrass, maintained a consistent seasonal dynamic. It showed early summer peaks and declines in late autumn across 2019–2024. The abrupt departure from this pattern from autumn 2025 onwards underlines the importance of within-year frequent observations. Annual surveys alone would have failed to identify the season in which collapse occurred and would have revealed the reduction too late.

Furthermore, in tidal flats—where shooting conditions, including tides, vary significantly over time—the ability to select the best images from a large number of shots is a major advantage. Even with low spatial resolution, it is possible to detect changes by conducting comparisons during specific seasons—such as early summer and winter, or early autumn when vegetation begins to decline—to assess differences from eelgrass and green algae, which become more prevalent during the summer.

Regarding seasonal variations before the rapid change in 2025, in other regions such as the Iwakuni area in Hiroshima Bay, the decline of eelgrass beds due to water temperature was already recorded after 2015, and previous studies have pointed out that eelgrass beds declined at temperatures exceeding 28°C (Abe et al., 2008; Hiraoka et al., 2000). Therefore, seasonal variations in the area may also be taken as an early warning signal of the decline of eelgrass beds. Accordingly, early-warning seasonal indicators that take into account seasonality and resilience are necessary.

### 4.3 Potential use in TNFD framework and EOVs of MBON

From an observational network perspective, "Monitoring Site 1000" for seagrass corresponds to the MBON EOVs framework (Duffy et al., 2025), providing a nationwide monitoring infrastructure in Japan. However, this is insufficient to capture the rapid collapse that occurred in less than a year. Although remote sensing has limitations in deeper areas, the number of sites where seasonal variations can be detected through high-frequency satellite observations should be increased for the EOV monitoring of coastal vegetation.

In recent years, there has been a growing demand not only for public monitoring but also for companies to disclose information regarding their efforts to mitigate the impact of their business activities and the effectiveness of their conservation initiatives. The Taskforce on Nature-related Financial Disclosures (TNFD) and corresponding directives such as the EU's CSRD (Corporate Sustainability Reporting Directive) are receiving particular international attention (Yamakita, 2025).

The State of Nature (SoN) metrics proposed by the Nature Positive Initiative (NPI), which underpin nature-related financial disclosures under the TNFD framework, are currently designed around a five-year assessment window. For highly dynamic systems such as seagrass beds, this approach carries a substantial baseline sensitivity problem: if the reference period coincides with a temporary peak (such as 1989 at the study site), subsequent decline appears disproportionately severe; if it coincides with a trough, the same decline may be underestimated. The seasonal variation documented here adds a further layer of complexity: comparisons across years are only meaningful when observations are made in the same season. For seagrass EOVs, we therefore recommend that (1) baselines be established over the longest available record and clearly justified ecologically, (2) seasonal standardisation be applied before inter-annual comparisons, and (3) years with extreme area anomalies be flagged or averaged out rather than being used as reference points.

The present workflow demonstrates the usefulness of affordable or freely available aerial imagery processed with open-source tools — including FOSS4G— for producing EOV-aligned time series. This approach is transferable to other APMBON member regions, particularly ASEAN countries, and to resource-limited coastal management contexts more broadly. Building on this foundation with OBIS Event Core-compatible outputs and BON in a Box integration represents a practical pathway toward scalable, low-cost seagrass monitoring aligned with international biodiversity frameworks.

Nevertheless, key preprocessing steps including orthorectification, geometric correction and labelling still depend largely on expensive proprietary software or manual adjustment. Furthermore, affordable imagery itself remains largely concentrated in Sentinel-2 archives accessible via Google Earth Engine and historical aerial photographs, which are difficult to obtain in many countries. Development of open-source tools that automate these preprocessing steps and consolidate multi-source imagery —including the growing volume of drone surveys— into analysis-ready databases would substantially lower the barriers to deploying this workflow globally, and is essential if open-source remote sensing is to serve as a routine component of seagrass EOV monitoring.

## 5. Conclusions

This study demonstrated that, by integrating historical aerial photographs, high-resolution satellite imagery, and satellite constellations using open-source analytical methods, it is possible to reconstruct both long-term and seasonal variations in seagrass vegetation with high accuracy, and to detect sudden ecosystem collapse. Both long-term and seasonal variations were significant, and while the recent disappearance of eelgrass fell within the range of these variations, it was qualitatively different; the event was likely influenced by elevated sea surface temperatures across a broad region, rather than by site-specific factors alone. For reporting frameworks such as TNFD and EOVs, using seagrass area as an indicator requires not only verification using sparse high-resolution images and annual field surveys but also the establishment of an appropriate time scale with high-frequency observations and standardized seasons.

## Acknowledgements

We express our sincere gratitude to the Ako Fisheries Cooperative Association for their cooperation during the field surveys; and to Marine Works Japan Ltd. for support with seagrass extraction and georectification. This study was partially supported by the Environmental Research and Technology Development Fund S23 (JPMEERF24S12306), 4-2302 (JPMEERF20234002), and S21 (JPMEERF23S12104); JST CREST (JPMJCR23J1); JSPS KAKENHI (20KK0246); and JST COI-NEXT (JPMJPF2206).

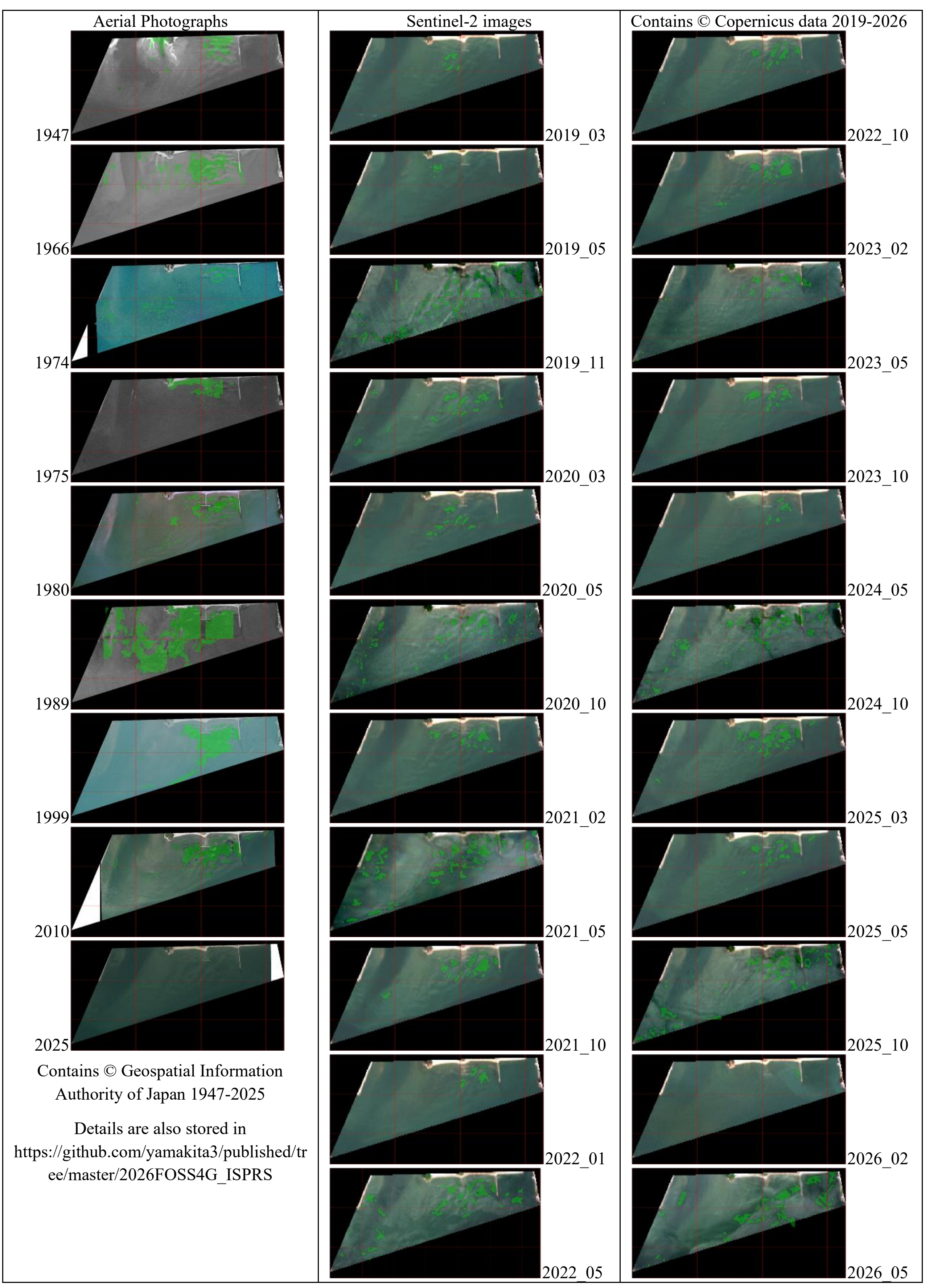

Appendix Figure S-1. Results of applying the seagrass extraction model. The left column shows the results of the analysis of grayscale (B/W) aerial photographs taken by GSI or US Army. The two right columns show the results for Sentinel-2 satellite imagery in winter, early summer, and autumn. Produced models and data are also shared in the GitHub repository linked above.